\documentclass[10pt,twocolumn,letterpaper]{article}

\usepackage{cvpr}              

\usepackage{graphicx}
\usepackage{amsmath}
\usepackage{amssymb}
\usepackage{booktabs}

\usepackage[pagebackref,breaklinks,colorlinks]{hyperref}

\usepackage[capitalize]{cleveref}
\crefname{section}{Sec.}{Secs.}
\Crefname{section}{Section}{Sections}
\Crefname{table}{Table}{Tables}
\crefname{table}{Tab.}{Tabs.}

\def\cvprPaperID{5141} 
\def\confName{CVPR}
\def\confYear{2022}

\begin{document}

\title{Video is All You Need: Attacking PPG-based Biometric Authentication}

\author{Lin Li\\
Swinburne University of Technology\\
{\tt\small linli@swin.edu.au}
\and
Chao Chen\\
James Cook University\\
{\tt\small chao.chen@jcu.edu.au}

\and
Lei Pan\\
Deakin University\\
{\tt\small l.pan@deakin.edu.au}

\and
Jun Zhang\\
Swinburne University of Technology\\
{\tt\small junzhang@swin.edu.au}

\and
Yang Xiang\\
Swinburne University of Technology\\
{\tt\small yxiang@swin.edu.au}
}

\maketitle

\begin{abstract}
Unobservable physiological signals enhance biometric authentication systems. 
Photoplethysmography (PPG) signals are convenient owning to its ease of measurement and are usually well protected against remote adversaries in authentication.
Any leaked PPG signals help adversaries compromise the biometric authentication systems, and the advent of remote PPG (rPPG) enables adversaries to acquire PPG signals through restoration. 
While potentially dangerous, rPPG-based attacks are overlooked because existing methods require the victim's PPG signals.
This paper proposes a novel spoofing attack approach that uses the waveforms of rPPG signals extracted from video clips to fool the PPG-based biometric authentication. 
We develop a new PPG restoration model that does not require leaked PPG signals for adversarial attacks.
Test results on state-of-art PPG-based biometric authentication show that the signals recovered through rPPG pose a severe threat to PPG-based biometric authentication. 
\end{abstract}

\section{Introduction}
\label{intro}

The security of biometric systems is enhanced with the adoption of physiological signals (like Electrocardiogram \cite{huang2021robust}, Electroencephalogram \cite{wang2020brainprint}, Photoplethysmography \cite{Hwang9130730}), assuming that attackers cannot easily obtain the victim's physiological signals \cite{Rathore3410158}. 
Photoplethysmography (PPG) signals are commonly measured by wearable devices integrated with cost-effective PPG signal sensors. 
For example, Apple Watch and Samsung Galaxy Fit2 use PPG sensors to monitor the heart health status.
PPG signal satisfies the basic properties (Universality, Distinctiveness, Permanence, Collectability \cite{jain2004introduction}) of the biometrics. 
PPG signals collected from wearable devices are invisible to the naked eye, which facilitates unnoticeable authentication or even continuous authentication. 

There are various studies on PPG-based user authentication relying on temporal features \cite{Zhao9155526}, spectrum features \cite{Donida978}, automated feature extraction through CNN-LSTM models \cite{Hwang9130730}, or many alike.
Moreover, industry vendors are investigating new PPG-based biometric authentication. 
Such authentication is designed to improve the security of existing authentication systems.
Some well-known patents are filed by Samsung \cite{jain2021real} and Nymi \cite{oung2020live}. 
These proposals use physiological signals, especially PPG, as unique user identifiers.
PPG-based biometric authentication is an emerging and promising research topic. 

Nevertheless, the unobservable characteristics of PPG signals are challenged by remote acquisition. 
The remote acquisition allows PPG signals to be obtained beyond the close contact physical distance, implying that the PPG-based biometric system loses its unobservable advantage. 
Several remote PPG (rPPG) methods are proposed to infer biometric signals, including discerning facial video sequences \cite{Huang9205299}, detecting arterial blood \cite{mocco2018new}, and monitoring heart rate information \cite{Seepers7892894,Calleja3}. 
Though some studies are provided for exploiting PPG signals \cite{Seepers7892894,Calleja3, karimian2019attack}, there is no attack targeting biometric authentication systems using rPPG. 
One reason is the overlook of signal features, for instance, the morphological features of the signal waveform are disregarded  \cite{Seepers7892894,Calleja3}. 
The second reason is the narrow view of the features of a single cardiac cycle used as a unique identifier for PPG-based biometric systems \cite{zhao2020trueheart,hwang2021variation}.
The third reason is the requirement of close physical contact between the attacker and the victim to obtain the high-quality PPG signals \cite{karimian2019attack}. 
Exploring the potentials of rPPG-based attacks is essential to assess the security of PPG-based biometric systems, which motivates this work.

\begin{figure*}[!ht]
  \centering
  \includegraphics[width=\linewidth]{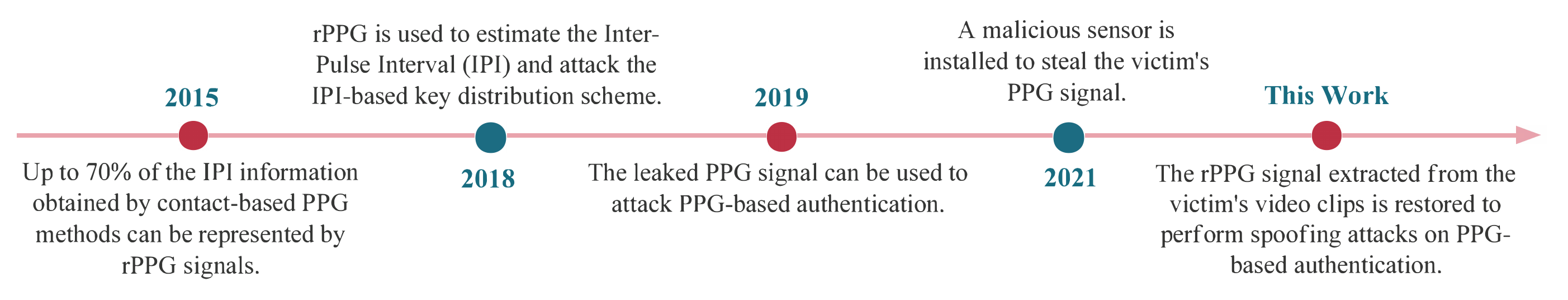}
  \caption{An evolving timeline for attacks against PPG-based biometric authentication.}
  \label{timeline}
\end{figure*}

In this paper, we propose a spoofing attack with video clips against PPG-based biometric authentication.
Our spoofing attack directly uses the rPPG signals extracted from a human face presented in videos. 
The technical challenge is the differences between the rPPG signal and PPG signal in the timing of signal attainment.  
To mitigate the signal difference, our method uses a generative model to restore the rPPG signal to the PPG signal.
The PPG signal is collected from the subject's fingertip using an FDA-approved oximeter.
As shown in~\cref{timeline}, generalizing attacks with fewer assumptions, this work successfully attacks PPG-based biometric authentication with nothing except an HD video clip of the victim's face. 
To the best of our knowledge, this is the first work to exploit the morphological features of rPPG to spoof PPG-based authentication.
The major contributions of this paper are:

\begin{itemize}
  \item We are the first to conduct a  spoofing attack on PPG-based biometric authentication using video clips. We propose a novel signal restoration method called SigR to accurately restore rPPG signals to PPG signals. 

  \item We model and measure the effectiveness of restoring PPG signals from rPPG signals extracted from video clips in the UBFC-PHYS \cite{Meziatisabour9346017} dataset, where human subjects are in different states  (`resting', `talking', or `calculating').
  
  \item We conduct experiments on the impact of various video quality factors (frame size, frame rate, bit-rate, and beauty filter) by comparing the results obtained on the spoofing attack.
\end{itemize}


\section{Related Work}
\label{relatedwork}

\subsection{PPG-based Biometric Authentication}
\begin{figure}[!ht]
  \centering
  \includegraphics[width=\linewidth]{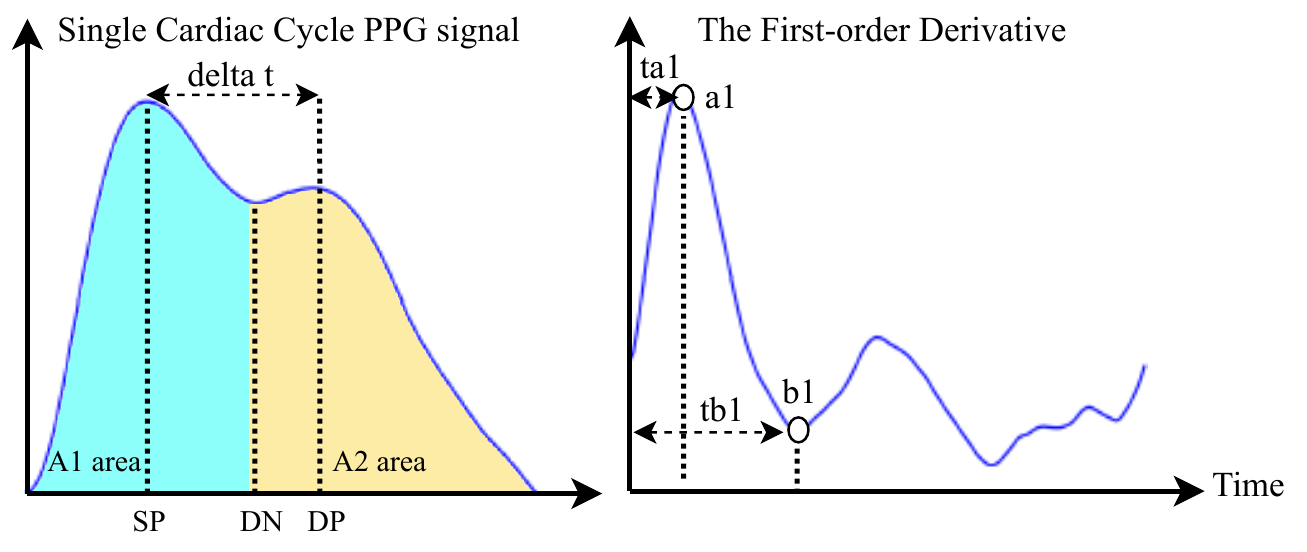}
  \caption{An example of fiducial point of PPG signal and the first-order derivative for a cardiac cycle. SP: systolic peak index, DN: dicrotic notch index, DP: diastolic peak index.}
  \label{wave}
\end{figure}

The PPG signal corresponds to a wealth of heart-related information distinct to the individual. 
Features from the PPG signal can be used for biometric applications \cite{cheng2019novel}.
Early studies extracted temporal features from the fiducial points of the PPG waveform, such as peak number, time interval, and upward/downward slope~\cite{Gu1222403}. 
The first and second-order derivative of the PPG signal serves as a useful feature for biometric authentication~\cite{RESITKAVSAOGLU20141}. 
Some features that are commonly used for biometric authentication are shown in~\cref{wave}.
Temporal features of the PPG signals are sensitive to noises, including baseline wander, motion artifact, and respiration \cite{karimian2017human}. 
To improve the robustness against noise, frequency-based features are obtained by applying transform methods to the PPG signal like Fourier transform \cite{donida2021biometric} and wavelet transform \cite{Yadav8411233}. 
Recent state-of-the-art PPG-based biometric authentication uses deep learning to learn features automatically from the raw data \cite{Biswas8607019, Hwang9130730}. 

\begin{figure*}[!ht]
  \centering
  \includegraphics[width=0.9\linewidth]{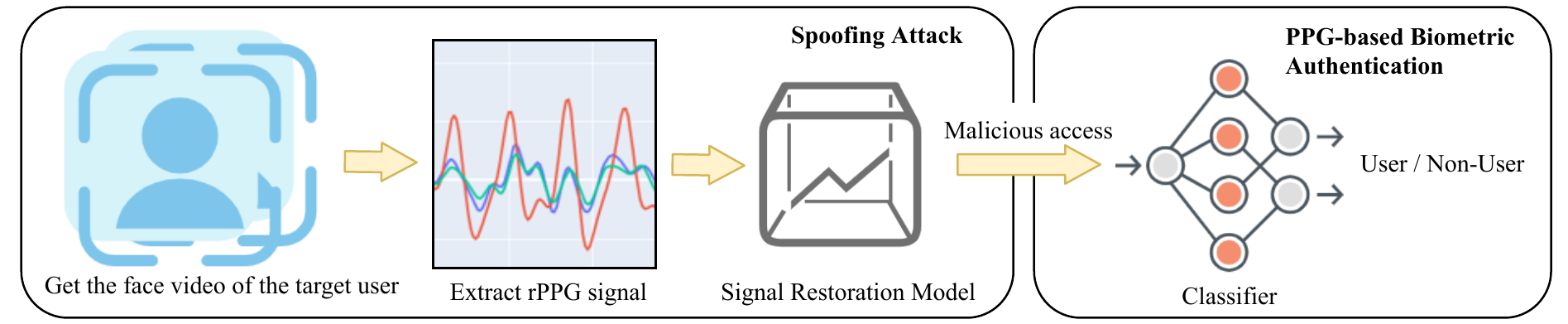}
  \caption{An attack flow of PPG-based biometric authentication. During biometric authentication, the PPG signal is captured by a contact device for signal processing, including noise filtering, heartbeat cycle separation, and signal normalization. Then the biometric authentication component distinguishes users and non-users. During the attack, the rPPG signal is obtained from the video frame containing the victim's face. The rPPG signal is restored to a PPG signal using the restoration model before the spoofing attack is finally performed on the classifier with the restored signal.}
  \label{workflow}
\end{figure*}

\subsection{Existing Attacks}
\label{remoteppg}
The emergence of rPPG has exposed PPG-based security systems to an unprecedented threat.
Up to 70\% of the inter-pulse interval (IPI) information obtained by contact-based PPG methods can be leaked by rPPG signals \cite{Calleja3}. 
IPI measures the time difference between two consecutive systolic peaks.
The IPI information is used as a random sequence in a key distribution scheme \cite{zhang2021h2k}. 
An attacker can use rPPG to mimic PPG for generating IPI-based biometric identifiers before compromising an IPI-based key distribution \cite{Seepers7892894}. 

Karimian et al.~\cite{karimian2019attack} use leaked signals to present spoofing attacks against PPG-based biometric authentication systems.
Three differential equations are used to synthesize a PPG signal for spoofing attacks. 
Then, two Gaussian functions are applied to convert the model parameters of an arbitrary PPG signal to the victim's to spoof the authentication system. 
Since PPG signals can be easily acquired from many measurement points on the human body, malicious PPG sensors attached to the body covertly help steal the victim's PPG signals.
Subsequently, the attacker can reproduce the victim's PPG signal based on the malicious sensor's recorded signal through a waveform generator \cite{hinatsu2021basic}.

However, these attacks require the victim's PPG signal that is challenging to acquire through close physical contact without the victim's awareness.
The assumption of the victim's PPG signals strongly limits the power of existing attacks. 
This limitation motivates this work to establish novel attacks with easily acquired videos. 
To our best knowledge, no existing attacks target PPG-based authentication systems by using video clips.

\subsection{Remote Photoplethysmography (rPPG)}
\label{Remoteppg}
To obtain rPPG signals from a video clip, anyone can calculate differences among the facial color channels.
The green channel has the strongest plethysmographic signal, corresponding to the absorption peak of hemoglobin \cite{verkruysse2008remote}. 
Principal Component Analysis (PCA) and Independent Component Analysis (ICA) are used to separate the pulse signal from the RGB channels \cite{poh2010non,lewandowska2011measuring}.
CHROM~\cite{de2013robust} projects the normalized RGB values to two orthogonal chromaticity vectors before calculating their difference to extract the PPG signal. 
The rotation angle between the skin pixel subspaces in the video frame is used to recover the rPPG signal \cite{wang2015novel}.
As a mature deep learning technique, CNN has been proposed for end-to-end video-based heart rate measurement \cite{chen2018deepphys}. 
To acquire a reliable rPPG feature representation, rPPGNet combines the skin segmentation task with the rPPG recovery task through a spatial-temporal convolutional network along with a skin-based attention module  \cite{yu2019remote}. 
GAN has been used in \cite{song2021pulsegan, lu2021dual} to improve the quality of continuous waveforms of rPPG signals. 
However, most existing works extract the heart rate without emphasizing the waveform's accuracy. 


\section{Our Spoofing Attack}
\label{threatmodel}
We aim to design a spoofing attack on PPG-based biometric authentication only by video clips. 
\cref{workflow} presents the workflow of our attack. 
\cref{assumption} introduces the threat model of our attack. 
Initially, we obtain the region of interest in a video clip through face detection. 
To acquire a suitable rPPG signal, we adopt CHROM (\cref{signalacquisition}). 
Finally, we elaborate on how to restore a rPPG signal to the PPG signal (\cref{signaltranslation}).

\subsection{Threat Model}
\label{assumption}
We outline the adversary's capabilities to understand the requirements of executing an attack against PPG-based biometric authentication systems.
In this paper, we assume that the attacker has access to videos containing the victim's face. 
Our assumption is practical and realistic because the rPPG signal from video extraction is free of physical contact with the victim. 
An active attacker may install a hidden HD camera to record the victim's face video to steal rPPG signals.
Conversely, passive attackers can retrieve videos shared by the victims through online social media platforms like YouTube, Facebook, and TikTok.
It makes the attack more stealthy and unnoticeable to the victim than any other existing method.
In previous work, attacks against PPG-based authentication almost always require the availability of the victim's leaked PPG signals. 
Our attack method does not require any leaked PPG signals, resulting in realistic attacks.

\begin{figure*}[!ht]
  \centering
  \includegraphics[width=0.95\linewidth]{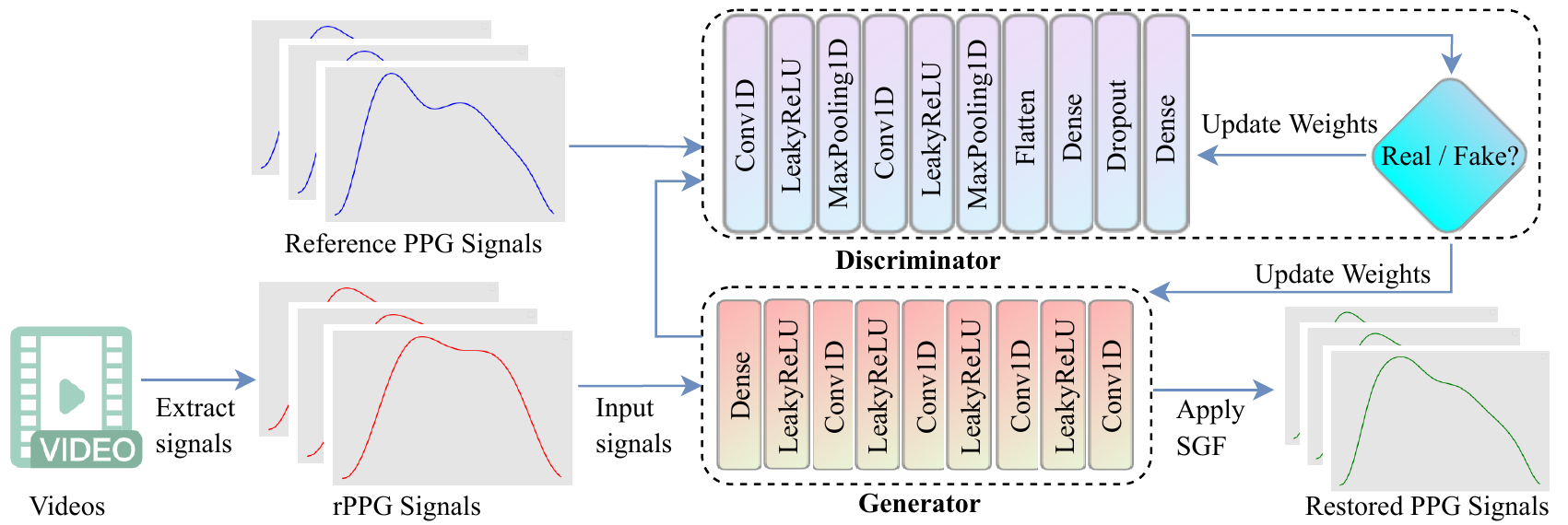}
  \caption{PPG signal restoration model structure. The generator produces the signal for PPG signal by feeding the rPPG signal. The discriminator is used to judge whether the generated signal belongs to the PPG signal. SGF:  Savitzky-Golay filter.}
  \label{gan}
\end{figure*}

\subsection{rPPG Acquisition}
\label{signalacquisition}
Our rPPG acquisition method consists of two steps.
In Step one, we mark the region of interest (ROI) in the video by segmenting the face skin area.
In Step two, we use the CHROM tool to extract the rPPG signal from the skin region of the human faces in the continuous video frames.

First, we detect the ROI in the video by using MTCNN \cite{zhang2016joint} to separate human faces from video frames.
By using MTCNN, we obtain the face positions at different granularity levels with accurate ROIs.
Having obtained the position of the face, we filter the ROI by the adaptive skin detection algorithm \cite{kolkur2016human}. 
By setting an appropriate threshold value, we exclude the pixels from non-skin sections, like background and hair, to mitigate the impact of background noise.
Now, we capture the skin region from the video clip.

Next, we use the chrominance-based method CHROM to extract the rPPG signal from the skin region. 
As CHROM is based on a skin optical reflection model, we can quickly acquire the signal with good robustness to motion artifacts. 
And CHROM works well on high-resolution face images alone. 
Currently, there are limited publicly available data samples that include the face video and corresponding PPG signals. 
Hence, we choose the CHROM method to extract the rPPG signal from the isolated ROIs.


\subsection{SigR: Our Restoration Model}
\label{signaltranslation}
After obtaining the rPPG signal, we construct a restoration model SigR to restore the rPPG signal to the PPG signal.
The PPG signals reflect changes in blood flow from facial skin and fingers.
However, we observe that the collected PPG and rPPG signals significantly differ because they emit at different distances from the heart and tissues. 
For instance, the difference in the signal's arrival time causes differences in the signal phase.
The difference in human tissues results in a variation in the magnitude of the waveform. 
Traditional methods struggle with modeling this relationship due to the human body's complex nature and the overwhelming noise in the external environment.

In this work, we propose a signal restoration model ($R$: rPPG  $\rightarrow$ PPG). 
It aims to learn the distribution of differences between signals from a small amount of data. 
It adapts GAN's network structure for signal processing.
Specifically, our generator $G$ takes the rPPG signal as the input to learn the latent space, approximating the generated signal close to the victim's PPG signal that spoofs the discriminator. 
As illustrated in~\cref{gan}, our generator $G$ has four Conv1D layers to capture the signal differences between one-dimensional signals.
LeakyReLU is used as the activation function.
The discriminator consists of two layers of Conv1D and MaxPooling1D layers, respectively. 
The discriminator checks the reference PPG signal and determines whether the restored signal output by the generator is acceptable or not.
We apply Wasserstein distance and gradient penalties to stabilize the training process~\cite{gulrajani2017improved}.


To remain consistent with existing work, the PPG signal of one cardiac cycle is used as a unique identifier for each user. 
We use one cardiac cycle of PPG signal and rPPG signal to train the GAN network.
To isolate individual cardiac cycle, the heartbeat segmentation is performed on the continuous signal with the beat separation algorithm in \cite{Lovisotto9150630}.
Segmented rPPG signals are fed into the generator for supervised latent space learning. 

Through adversarial learning between the generator and the discriminator, the generator's output signal yields a distribution similar to the PPG signal. 
As shown in~\cref{gan}, our restoration model combines the trained generator and the subsequent filter.
We apply the Savitzky-Golay filter~\cite{schafer2011savitzky} to smooth the signal. 
After restoring the single cardiac cycle rPPG signal to the PPG signal, we average the multiple restored signals to reduce the  bias.

\section{Experiment}
\label{experiment}
In this section, we evaluate the effectiveness of the proposed spoofing attack method.
We conduct a series of experiments on the UBFC-PHYS \cite{Meziatisabour9346017} dataset, including the use of rPPG signals in three different states (`resting', `talking', or `calculating'). 
We also evaluate the attack success rate at different video qualities, including frame rate, frame size, bit-rate, and use of beauty filter.

\subsection{Datasets}

\textbf{UBFC-PHYS \cite{Meziatisabour9346017}:} 
It includes facial videos of 56 participants and their PPG signals.
Each participant has a three-minute video with different states.
The video resolution is $1024 \times 1024$ with a frame rate of 35 FPS and 227,474 Kbps bit-rate. 
The sampling rate of the PPG signal is 65 HZ. 
The data was collected under three different states --- `resting', `talking', or `calculating'.  
The `resting' state requires the participants to remain relaxed and silent; the `talking' state demands the participants to talk with a facilitator. Moreover, the `calculating' state challenges the participants with mathematical tasks and requires them to read their answers aloud. 
These conditions simulate environments with different pressure levels of a human. 
Due to the participants' physical movement, faces are not recognizable in some video frames, resulting in signal interruptions.
Therefore, we exclude the videos with signal interruptions. 
In the end, we acquire 51 video data in the `resting' state video, 40 videos in the `talking' state, and 47 videos in the `calculating' state. 
In our experiments, each user is treated as a victim separately, while the other users are considered non-victims.
For each video clip, we extract the rPPG signals in a range between 150 and 287 cardiac cycles, which corresponds to a video clip of 3 minutes long.  
All non-victim users have PPG signals embedded in approximately  11,800 cardiac cycles. 


\begin{table*}[!ht]
\small
\centering
\begin{tabular}{l c c c c c c c c c c c c c c}
\toprule
\textbf{Method}& \textbf{SP} &	\textbf{DN}	 & \textbf{DP} & \textbf{A2\_area} &	\textbf{A1\_area} & \textbf{A2\_A1\_ratio}	& \textbf{a1} &	\textbf{b1}  & \textbf{ta1} &	\textbf{delta\_t} \\ \midrule
rPPG &	0.3107& 0.3264& 0.2817& 0.3233& 0.2802& 0.3654& 0.2623& 0.3592& \textbf{0.3084}& 0.2347\\ \midrule
GMM& 0.5019&  0.4180&  0.4866&  0.4290&  0.4786&  0.3929&  0.6251&  0.4649& 0.6051& 0.4027 \\ \midrule
GP &	0.2849& 0.3068& 0.2564& 0.3178& 0.2525& 0.3770& \textbf{0.2525}& 0.2882 & 0.3892 & 0.2802\\ \midrule
SigR  & \textbf{0.2370}& \textbf{0.2498}& \textbf{0.2145}& \textbf{0.2549}& \textbf{0.2164}& \textbf{0.3141}& 0.3849& \textbf{0.2627}& 0.3319& \textbf{0.2217}\\ \bottomrule
\end{tabular}
\caption{Average Kolmogorov-Smirnov test of features between the restored PPG signal and the reference PPG signal collected in the `resting' state. SP: systolic peak index, DN: dicrotic notch index, DP: diastolic peak index. }
\label{examkl}
\end{table*}

\begin{figure*}[!ht]
  \centering
  \includegraphics[width=0.9\linewidth]{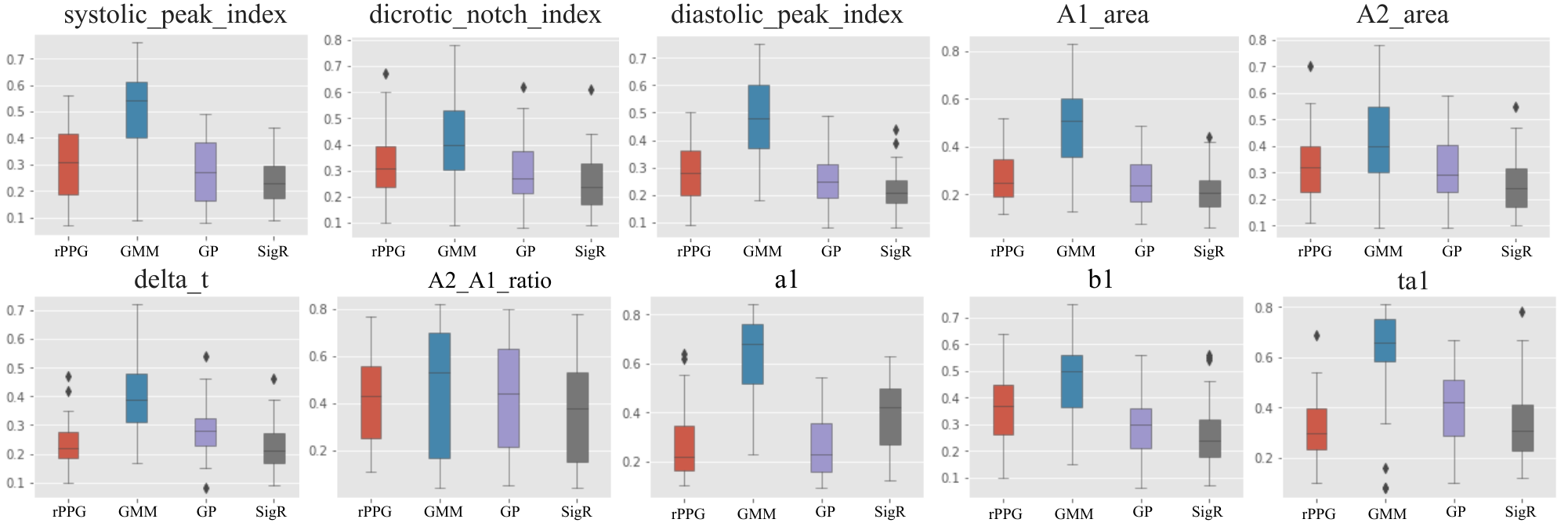}
  \caption{Comparisons of the Kolmogorov-Smirnov test for different restored PPG signals with the reference PPG signal. Each plot reports the comparisons of the four signals with respect to a specific feature. }
  \label{featurecompare}
\end{figure*}

\subsection{Experimental Settings}
We use Keras\footnote{\url{https://www.tensorflow.org/}} to simulate a state-of-the-art biometric authentication system with a CNN/LSTM network \cite{Hwang9130730}.
We adjust the average equal error rate (EER) of the biometric authentication component to approximately 14\%, consistent with the EER used in \cite{Hwang9130730}.
Our model's architecture resembles a state-of-the-art PPG-based biometric authentication's architecture \cite{Hwang9130730}.
We use the open-source framework pyVHR \cite{Boccignone9272290} to extract the rPPG signals from video clips. 

We label the victim's signal as category 1 and other users' signal as category 0. 
For each user in the dataset, we use all the contact data of the victim and one-tenth of other users' as the training data. 
The other users' data is used as a control group to evaluate our spoofing attacks. 
For the restoration model, we use the rPPG signal as the input to the generative model and the PPG signal as the discriminator's reference.
To prevent the model from learning the variations between the victim's rPPG and PPG, we exclude the victim's data when training the restoration model for each victim. 

In our experiments, EER is used to evaluate the performance of the biometric system. 
EER is the error rate where the false acceptance rate (FAR) and the false rejection rate (FRR) are equal. 
FAR and FRR are the most commonly used indicators in biometric systems. 
FAR indicates the possibility that the system incorrectly accepts access by an unauthorized user. 
FRR represents the possibility of the system denying access to an authorized user. 
For evaluating the spoofing attack, we use FAR as an indicator.
Because we need to evaluate whether the system incorrectly accepts restored PPG signals injected by the attacker, the higher FAR indicates that the restored PPG signals are more threatening to the authentication system.
While SigR is proposed for signal restoration using GAN, the component of signal restoration in the new spoofing attack can be implemented by other machine learning/deep learning models.
As a comparison, we use the Gaussian process (GP) and Gaussian mixture model (GMM) as baseline models. 
We use the Stochastic Variational Gaussian Process~\cite{hensman2013gaussian} in GPflow~\cite{GPflow2017} as the GP implementation, and GaussianMixture from scikit-learn~\cite{pedregosa2011scikit} as the GMM implementation.

\subsection{Experimental Results}
\label{result}

\begin{figure}[!ht]
  \centering
  \includegraphics[width=\linewidth]{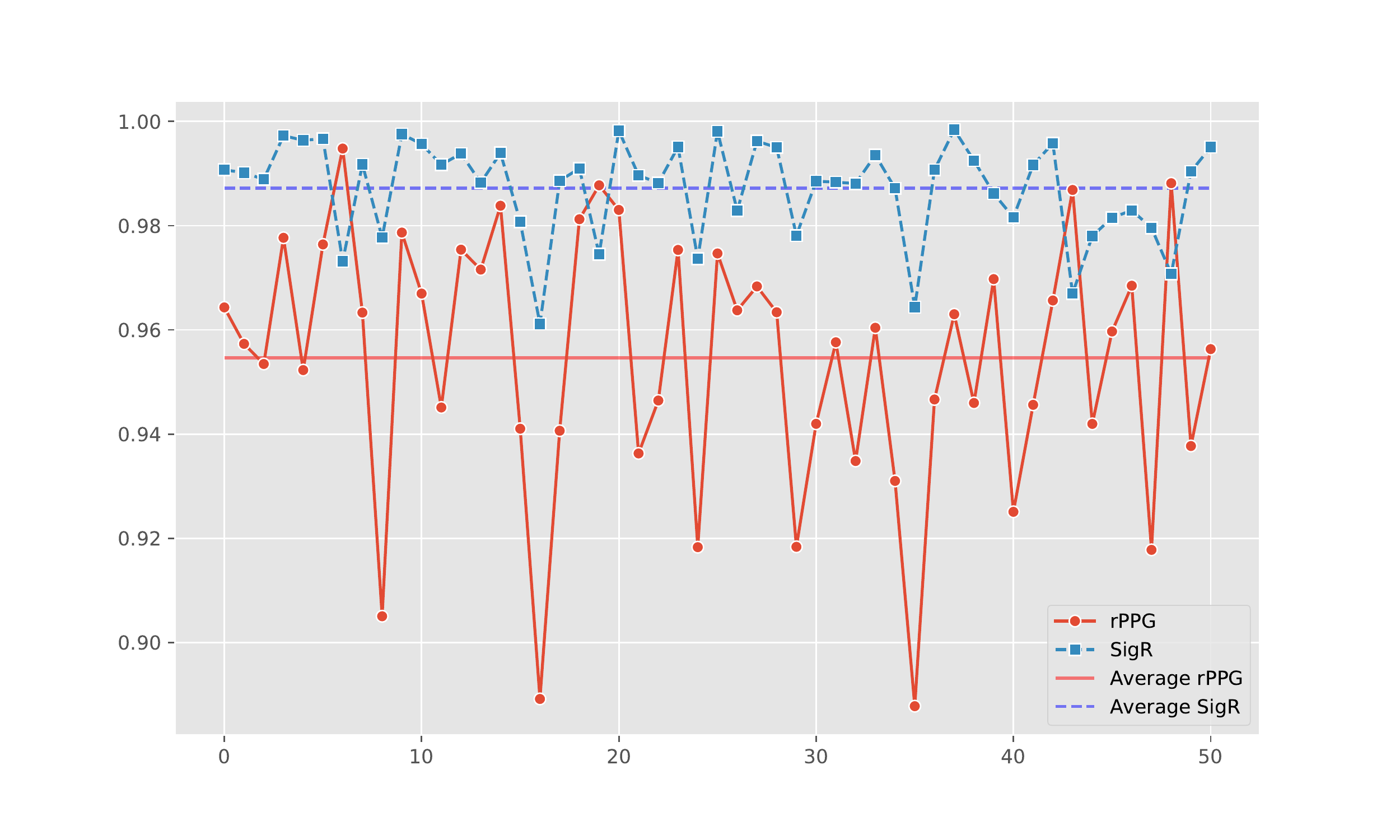}
  \caption{Pearson correlation coefficients of the heartbeat waveforms of the rPPG harvested from the videos and the PPG signal restored by SigR.}
  \label{pearson}
\end{figure}

\subsubsection{PPG Restoration Performance}
In our attacking process, the more similar the restored signal is to the PPG signal, the higher the signal quality is. 
Since the Kolmogorov-Smirnov test (KS) is sensitive to the probability distributions' location and shape, we use it to measure the similarity of two signals.
The smaller the value of the KS test, the more similar the distributions of the signals are.
\cref{examkl} shows the results of KS tests between rPPG, GMM, GP, SigR (our method), and PPG waveform features. 
These features are considered to be associated with individuals~\cite{Lovisotto9150630}.
The KS distribution of specific features is listed in~\cref{featurecompare}.
We observe that most of the feature distributions from SigR are closer than other methods to PPG. 

Herein, we compare the correlation between the signals. 
As shown in~\cref{pearson}, blue squares show the Pearson correlation coefficient between the PPG signal restored by SigR and the reference PPG signal. 
Red dots show the Pearson correlation coefficient between the original rPPG signal harvested from the video and the reference PPG signal.
We find that the coefficient between the rPPG signal harvested from the video and the reference PPG signal is drastically unstable. 
Conversely, the coefficient of the PPG signal restored by SigR and the reference PPG signal is steady between 0.96 and 1.00. 
Most of the signals restored by SigR have higher coefficients than directly harvested rPPG signals, indicating that the PPG signals restored by SigR are positively correlated with the reference PPG signal. 
Our restoration model successfully makes the restored PPG signal significantly closer to the reference PPG signal than the original rPPG signal harvested from the video.
The average increase in the correlation coefficient is 3.3\%, with a maximum increase of 7.7\%.

These results show that SigR well learns the relationship information between rPPG and PPG signals from the training data. 
SigR can use the learned information to restore a PPG signal from the corresponding rPPG signal that is the distorted or noise contaminated version of the PPG signal.

\begin{table}[t]
\small
\centering
\begin{tabular}{l c c c }
\toprule
\textbf{Status}  & \textbf{T1}  & \textbf{T2} & \textbf{T3}  \\\midrule
Random Attack & 0.14   & 0.15   & 0.15  \\ \midrule
MTTS-CAN  & 0.09   & -   & -  \\  \midrule
Victim rPPG Signal Attack  & 0.25   & 0.19   & 0.21   \\ \midrule
Victim GMM Signal Attack & 0.34   & 0.27  & 0.26   \\ \midrule
Victim GP Signal Attack & 0.34  & 0.27   & 0.28   \\ \midrule
Victim SigR Signal Attack & 0.40   & 0.27  & 0.29   \\ \midrule
Mean MTTS-CAN  & 0.13   & -   & -  \\ \midrule
Mean rPPG Signal Attack & 0.34   & 0.35   & 0.35   \\ \midrule
Mean GMM Signal Attack & 0.53   & 0.46   & 0.48  \\ \midrule
Mean GP Signal Attack & 0.43   & 0.44  & 0.40   \\ \midrule
Mean SigR Signal Attack & \textbf{0.61}   & \textbf{0.49}   & \textbf{0.57}   \\ \bottomrule
\end{tabular}
\caption{Spoofing attack FAR results of UBFC-Phys. T1: resting, T2: talking, T3: calculating. Random attack indicates the non-victim's PPG signal captured by the fingertip. Victim means a single cardiac cycle from the victim signal. rPPG, GMM, GP, SigR indicate methods of acquiring signals. The mean signal is the mean value of multiple signals. We use the mean value to maximize the FAR results for each victim.}
\label{result1}
\end{table}

\subsubsection{Attack Success Rates in Different States} 
The victim in the video clip may be in an arbitrary state (`resting', `talking', or `calculating').
\cref{result1} lists the FAR results of our method on the UBFC-Phys dataset.
Since PPG signals have distinct patterns related to human individuals, the quality of randomly generated data is insufficient to compromise the authentication system. 
Our random attacks use PPG signals collected from non-victim users as the input.
We observe that FAR of the random attack is 14\%, while it is 25\% of rPPG signal attack, which indicates that it is easier to directly apply the victim's rPPG signal than random attack in the `resting' state to mislead the biometric authentication system. 
Attacking the system using SigR's generated signal has a 15\% increase in FAR over the original rPPG signal, suggesting that our approach significantly increases the possibility of spoofing the authentication system.
A recently published rPPG extraction method named MTTS-CAN \cite{NEURIPS2020_e1228be4} only achieved 0.09 FAR result and 0.13 mean-treated FAR in the resting state. It performs even worse than random attacks. 
In addition, attacking using the mean-treated SigR signal performs the best (61\% in terms of FAR), increasing over 30\% in FAR compared to the original rPPG signal.  
It indicates that every two attempts will succeed in breaking through the authentication system. 
It makes our attack more realistic than the existing works as a real-world authentication system usually allows three attempts \cite{wang2021attacks}.  
We also find almost no difference in the FARs of the mean-treated rPPG signal for the three states, indicating that the mean-treatment reduces the influence of noise on the signal.

Compared to the rPPG signal in the `resting' state, the FARs in both `talking' and `calculating' states are reduced. 
`Talking' is usually accompanied by visible body movement, increasing the signal's noise. 
The `calculating' state simulates the person under pressure when the Autonomic Nervous System mediates the pressure.
Since the Autonomic Nervous System controls cardiac activity, it changes the PPG signals \cite{chauhan2018real}. 
Meanwhile, the restored PPG signals are also affected.


\subsubsection{Attack Success Rates in Different Video Qualities} 
To evaluate our attack FAR with different video qualities, we choose the video with the highest quality of the extracted rPPG signal from UBFC-PHYS as the benchmark. 
Through FFmpeg\footnote{FFmpeg is an open-source video transcoding tool used by many video transcoding software tools. It is available at \url{https://www.ffmpeg.org/}.}, we convert each original video clip into multiple video clips with different settings --- resolution ($1024 \times 1024$, $512 \times 512$, $256 \times 256$), frame rate ($35.14$, $30$, $20$ FPS), bit-rate ($227474$, $113865$, $255$ Kbps).
Videos with human faces are abundant on social networks. 
These videos are usually processed with beauty filters to achieve a nice visual effect. 
We use FFmpeg's `smartblur' filter to blur the facial area in input video without affecting the facial outline, which works similarly to the beauty filters.

\cref{comparequality} shows the authentication system's FAR results for the rPPG and restored signals at different video qualities. 
FAR drops significantly when the video's frame rate, frame size, or bit-rate decreases. 
Among all the parameters, the frame rate has the most significant influence on the rPPG signal result because the signal is extracted from each frame's color channel. 
A lower frame rate infers a lower sampling rate. 
A low sampling rate increases the difference between rPPG signal and PPG signal. 
The smartblur filter also significantly impacts the quality of the extracted rPPG signal.  
As smartblur uses a Gaussian filter to smooth the image and simultaneously alter the pixel differences between video frames.
We note that bit-rate slightly affects FAR, e.g., FAR is 0.62 when the bit-rate drops to 255 Kbps. 
When the bit-rate exceeds the threshold, the quality of the video image will be reduced.
Meanwhile, the FAR results for SigR signals are significantly less influenced by the video quality than the rPPG signal. 
When the video quality decreases from 35 FPS to 30 FPS, the average FAR of SigR's PPG drops by 5\%, but the FAR of the rPPG signal drops by 12\%. 
Regarding the frame rate that has the most significant impact on the rPPG signal, SigR shows good robustness as the FAR does not vary substantially. 
SigR learns a good representation of the latent space, making it insensitive to the rPPG signal sampling rate changes.
Resolution is the most influential of all factors on the FAR of SigR.
The lower resolution increases the contribution of each pixel in the video, which causes the random noise occurring in the signal we acquire to be more pronounced.
Regarding the bit-rate and smartblur filter, although SigR is affected, the impact is significantly smaller than that on rPPG.

\begin{table}
\small
\centering
\begin{tabular}{l c c}
\toprule
\textbf{Video Quality}      & \textbf{rPPG} & \textbf{SigR} \\ \midrule
Original Video (FR: 35 FPS)           & 0.75  &  \textbf{0.96}  \\ \midrule
FR: 30 FPS  & 0.63 & \textbf{0.91}   \\ \midrule
FR: 20 FPS  & 0.54 & \textbf{0.92} \\ \midrule
FS: $512 \times 512$ & 0.63 & \textbf{0.88}       \\ \midrule
FS: $256 \times 256$ & 0.60 & \textbf{0.81}       \\ \midrule
BR: 113,865 Kbps     & 0.71 & \textbf{0.93}   \\ \midrule
BR: 255 Kbps        & 0.62 & \textbf{0.92}  \\ \midrule
Filter: smartblur        & 0.64 & \textbf{0.92 } \\ \midrule
\end{tabular}
\caption{Comparisons of FARs in different video qualities. The parameters not specified in detail are consistent with the ones of the original video. FR: frame rate, FS: frame size, BR: bit-rate. }
\label{comparequality}
\end{table}

\begin{table*}
\centering
\begin{tabular}{l c c c c c c c c c c }
\toprule
\textbf{Video Quality} & \textbf{SP} &	\textbf{DN}	 & \textbf{DP} &  	\textbf{delta\_t} &	\textbf{A1\_area} & \textbf{A2\_area} & \textbf{A2\_A1\_ratio}	& \textbf{a1} &	\textbf{b1}  & \textbf{ta1} \\ \midrule
Original Video   &  0.13& 0.10& 0.09& 0.14& 0.18& 0.16& 0.47& 0.28& 0.44 & \textbf{0.28} \\ \midrule
FR: 30 FPS  & 0.16& 0.13& 0.15& 0.10& 0.21& 0.16& \textbf{0.61}& 0.19& 0.24& 0.15  \\ \midrule
FR: 20 FPS  & 0.12&  0.12&  \textbf{0.24}&  \textbf{0.20}&  0.20&  0.23&  0.56&  \textbf{0.39}&  \textbf{0.47}&  0.21 \\ \midrule
FS: $512 \times 512$ & 0.26&   \textbf{0.34}&   0.21&   0.11&   \textbf{0.37}&   \textbf{0.36}&   0.37&   0.15&   0.25&   0.11  \\ \midrule
FS: $256 \times 256$ & 0.26&  0.29&  0.23&  0.12&  0.36&  0.33&  0.22&  0.26&  0.13&  0.17 \\ \midrule
BR: 113,865 Kbps     & 0.18&  0.14&  0.17&  0.10&  0.23&  0.21&  0.46&  0.17&  0.38&  0.23 \\ \midrule
BR: 255 Kbps        & 0.22&  0.22&  0.14&  0.12&  0.26&  0.25&  0.51&  0.15&  0.32&  0.14 \\ \midrule
Filter: smartblur   & \textbf{0.27}&  0.22&  0.16&  0.06&  0.32&  0.26&  0.45&  0.20&  0.29&  0.09 \\ \midrule
\end{tabular}
\caption{Comparisons of the KS test results with rPPG signal and PPG signal features captured at the different video quality. FR: frame rate, FS: frame size, BR: bit-rate. The bold values indicate that the rPPG signal captured with the video parameter has the most significant difference from the PPG signal regarding this feature. SP: systolic peak index, DN: dicrotic notch index, DP: diastolic peak index. }
\label{comparequalityfeature}
\end{table*}

As shown in \cref{comparequalityfeature}, we compare the KS test results of the rPPG signal acquired at different video qualities with the original PPG signal. 
We find that the waveform of the captured rPPG signal varies with different video parameters. 
The videos (20 FPS) with the worst FAR results have the largest KS values in four features --- dicrotic notch, $a1$, $b1$, and delta $t$. 
The low frame rate affects on features of the original waveform and the first-order derivatives.
Frame size and the `smartblur' filter mainly affect the feature of the original waveform. 
We restore the PPG signal from rPPG, and the changes in the rPPG waveform also affect the restored signal.

\begin{figure}[t]
  \centering
  \includegraphics[width=\linewidth]{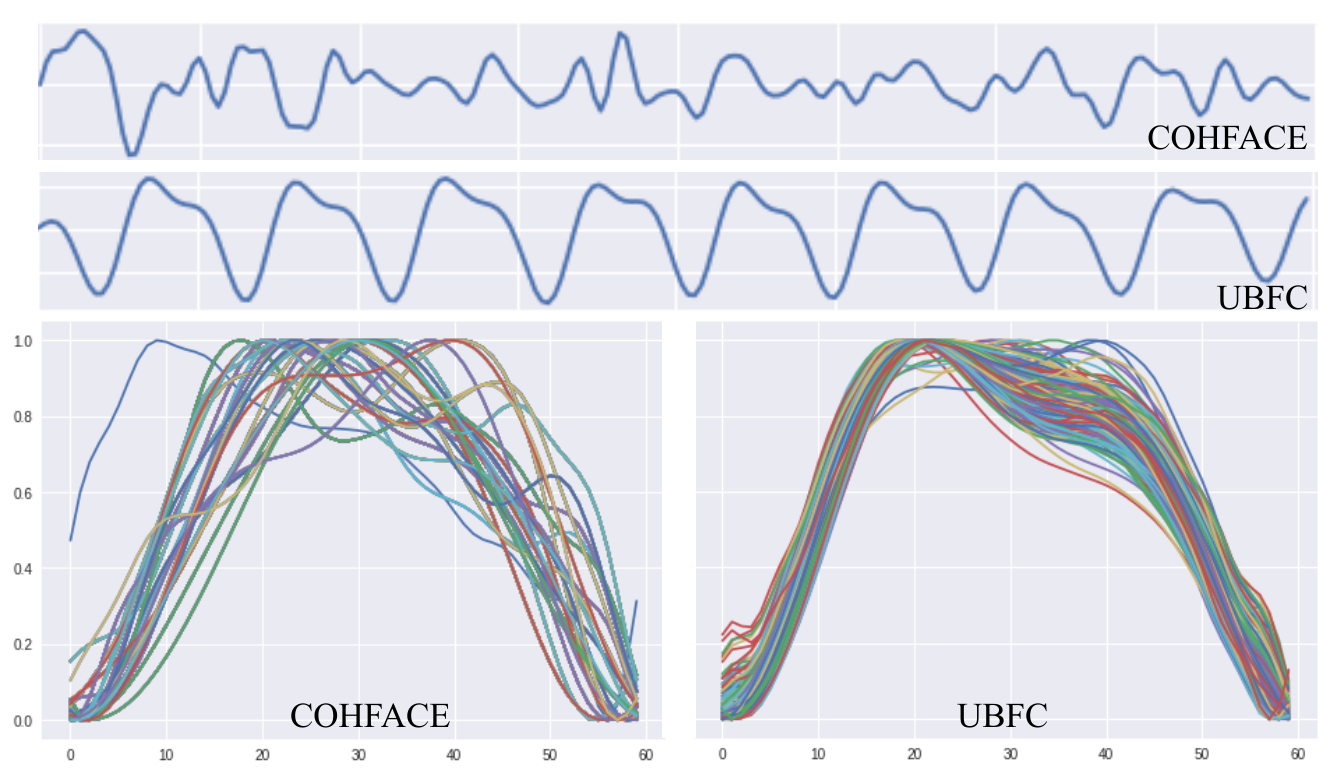}
  \caption{Comparisons of rPPG signals extracted from the UBFC-Phys and COHFACE datasets. The bottom left figure shows single cardiac cycle of a participant in COHFACE. The bottom right figure shows a UBFC-Phys sample with multiple cardiac cycles for a participant. }
  \label{cohfacecompare}
\end{figure}

\subsection{Discussion}
Our experimental results demonstrate the threat of rPPG to PPG-based biometric authentication, and our restoration model can significantly amplify the threat. 
Our proposed method restores rPPG to PPG signals that mislead the PPG-based biometric authentication setups. 
Our restoration model will strengthen with more high-quality datasets by eliminating the variances across datasets caused by different camera parameters, shooting distances, and lighting conditions.
Though our results are derived from the UBFC-PHYS dataset, our method has huge potentials for accurate PPG signal acquisition in telemedical scenarios. 
Due to this paper's scope, we will experiment with more videos captured under different conditions to improve the restoration model as future work. 

\noindent
\textbf{Defensive Strategies:} 
A common defense strategy against spoofing attacks may add an authentication component before activating the facial authentication system \cite{chingovska2014biometrics}.
The restriction of accessing high-quality video may decrease the success rate of our spoofing attack since the rPPG signal is affected by the video quality.
Furthermore, protecting the HD video of the potential victim is a key for defense because our method requires a high-quality video clip of the victim to isolate the rPPG signals for a single heartbeat cycle before performing the restoration. 
We conduct further experiments on the COHFACE dataset \cite{heusch2017reproducible}.
All rPPG signals recovered from COHFACE are insufficient to compromise the biometric authentication setup. 
As shown in~\cref{cohfacecompare}, the quality of the waveform obtained by the rPPG signal is low in the COHFACE dataset. 
Even the rPPG signal from one heartbeat cycle of the same individual varies drastically.
The frame rate of COHFACE videos is 20 FPS, the resolution is $640 \times 480$, and the bit-rate is 255 Kbps. 
These parameters are significantly lower than the parameters in the UBFC-PHYS dataset.
In the worst-case scenario, when the high-quality video needs to be released to the public, anyone may use some easily accessible means like beauty filters to mitigate leaking rPPG signals.

\section{Conclusion}
\label{conclusion}
As the shortcomings of traditional authentication solutions have become increasingly apparent, researchers have turned to explore new solutions like PPG-based authentication. 
However, we find that PPG-based biometric methods are vulnerable to spoofing attacks. 
To gain insights on the security of the PPG-based biometric authentication techniques, we propose a new spoofing attack requiring videos only.
The challenge lies in the subtle differences in the phase and shape between the face's rPPG signal and the actual PPG signal.
In this paper, we propose a signal restoration model to restore rPPG to PPG signals. 
For the PPG-based biometric authentication, our results confirm the severity of the attack. 
We experimented with rPPG signals from videos in different states, and the results were consistently and significantly higher than random attacks. 
Our generative model restores the rPPG signals well in different states. 
Our empirical studies achieved the highest average success rate of 0.62, indicating that rPPG poses a severe threat to PPG-based biometric authentication.
We also find that the quality of the video significantly affects the threat of the rPPG signal to the authentication system.
Our spoofing attack is still practical in real world scenarios. For example, popular social networking applications like Facebook (Meta), Youtube and Tiktok, host millions of videos in HD quality. Today, most smartphones are capable of shooting HD quality video.
Our empirical study shows that high-quality rPPG signals can be extracted from such videos. 
To mitigate such spoof attacks, we recommend alleviating rPPG signal leakage by adding a beauty filter before releasing HD videos.

{\small
\bibliographystyle{ieee_fullname}

\begin{thebibliography}{10}\itemsep=-1pt

\bibitem{Biswas8607019}
Dwaipayan Biswas, Luke Everson, Muqing Liu, Madhuri Panwar, Bram-Ernst Verhoef,
  Shrishail Patki, Chris~H. Kim, Amit Acharyya, Chris Van~Hoof, Mario
  Konijnenburg, and Nick Van~Helleputte.
\newblock Cornet: Deep learning framework for ppg-based heart rate estimation
  and biometric identification in ambulant environment.
\newblock {\em IEEE Transactions on Biomedical Circuits and Systems},
  13(2):282--291, 2019.

\bibitem{Boccignone9272290}
Giuseppe Boccignone, Donatello Conte, Vittorio Cuculo, Alessandro D’Amelio,
  Giuliano Grossi, and Raffaella Lanzarotti.
\newblock An open framework for remote-ppg methods and their assessment.
\newblock {\em IEEE Access}, 8:216083--216103, 2020.

\bibitem{Calleja3}
Alejandro Calleja, Pedro Peris-Lopez, and Juan~E. Tapiador.
\newblock Electrical heart signals can be monitored from the moon: Security
  implications for ipi-based protocols.
\newblock In Raja~Naeem Akram and Sushil Jajodia, editors, {\em Proceedings of
  the Information Security Theory and Practice}, volume 9311, pages 36--51,
  Cham, 2015. Springer International Publishing.

\bibitem{chauhan2018real}
Utkarsh Chauhan, Norbert Reithinger, and John~R Mackey.
\newblock Real-time stress assessment through ppg sensor for vr biofeedback.
\newblock In {\em Proceedings of the 20th International Conference on
  Multimodal Interaction: Adjunct}, pages 1--5, 2018.

\bibitem{chen2018deepphys}
Weixuan Chen and Daniel McDuff.
\newblock Deepphys: Video-based physiological measurement using convolutional
  attention networks.
\newblock In {\em Proceedings of the European Conference on Computer Vision
  (ECCV)}, pages 349--365, 2018.

\bibitem{cheng2019novel}
Shuo Cheng, Yongxin Chou, Jicheng Liu, Ya Gu, and Xufeng Huang.
\newblock A novel identity authentication method by modeling
  photoplethysmograph waveform.
\newblock In {\em Proceedings of the International Conference on Control,
  Automation and Information Sciences (ICCAIS)}, pages 1--5. IEEE, 2019.

\bibitem{chingovska2014biometrics}
Ivana Chingovska, Andre~Rabello Dos~Anjos, and Sebastien Marcel.
\newblock Biometrics evaluation under spoofing attacks.
\newblock {\em IEEE Transactions on Information Forensics and Security},
  9(12):2264--2276, 2014.

\bibitem{de2013robust}
Gerard De~Haan and Vincent Jeanne.
\newblock Robust pulse rate from chrominance-based rppg.
\newblock {\em IEEE Transactions on Biomedical Engineering}, 60(10):2878--2886,
  2013.

\bibitem{Donida978}
Ruggero Donida~Labati, Vincenzo Piuri, Francesco Rundo, Fabio Scotti, and
  Concetto Spampinato.
\newblock Biometric recognition of ppg cardiac signals using transformed
  spectrogram images.
\newblock In {\em Proceedings of the ICPR International Workshops and
  Challenges on Pattern Recognition}, pages 244--257, Cham, 2021. Springer
  International Publishing.

\bibitem{donida2021biometric}
Ruggero Donida~Labati, Vincenzo Piuri, Francesco Rundo, Fabio Scotti, and
  Concetto Spampinato.
\newblock Biometric recognition of ppg cardiac signals using transformed
  spectrogram images.
\newblock In {\em Proceedings of the ICPR Workshop on Mobile and Wearable
  Biometrics (WMWB)}, volume 12668, pages 244--257. Springer, 2021.

\bibitem{Gu1222403}
Y.Y. Gu, Y. Zhang, and Y.T. Zhang.
\newblock A novel biometric approach in human verification by
  photoplethysmographic signals.
\newblock In {\em Proceedings of the 4th International IEEE EMBS Special Topic
  Conference on Information Technology Applications in Biomedicine}, pages
  13--14, 2003.

\bibitem{gulrajani2017improved}
Ishaan Gulrajani, Faruk Ahmed, Martin Arjovsky, Vincent Dumoulin, and Aaron
  Courville.
\newblock Improved training of wasserstein gans.
\newblock In {\em Proceedings of the 31st International Conference on Neural
  Information Processing Systems}, pages 5769--5779, 2017.

\bibitem{hensman2013gaussian}
James Hensman, Nicol{\`o} Fusi, and Neil~D Lawrence.
\newblock Gaussian processes for big data.
\newblock In {\em Proceedings of the 29th Conference on Uncertainty in
  Artificial Intelligence}, pages 282--290, 2013.

\bibitem{heusch2017reproducible}
Guillaume Heusch, Andr{\'e} Anjos, and S{\'e}bastien Marcel.
\newblock A reproducible study on remote heart rate measurement.
\newblock {\em arXiv preprint arXiv:1709.00962}, 2017.

\bibitem{hinatsu2021basic}
Shun Hinatsu, Daisuke Suzuki, Hiroki Ishizuka, Sei Ikeda, and Osamu Oshiro.
\newblock Basic study on presentation attacks against biometric authentication
  using photoplethysmogram.
\newblock {\em Advanced Biomedical Engineering}, 10:101--112, 2021.

\bibitem{Huang9205299}
Po-Wei Huang, Bing-Jhang Wu, and Bing-Fei Wu.
\newblock A heart rate monitoring framework for real-world drivers using remote
  photoplethysmography.
\newblock {\em IEEE Journal of Biomedical and Health Informatics},
  25(5):1397--1408, 2021.

\bibitem{huang2021robust}
Yuwen Huang, Gongping Yang, Kuikui Wang, Haiying Liu, and Yilong Yin.
\newblock Robust multi-feature collective non-negative matrix factorization for
  ecg biometrics.
\newblock {\em Pattern Recognition}, 123:108376, 2022.

\bibitem{hwang2021variation}
Dae~Yon Hwang, Bilal Taha, and Dimitrios Hatzinakos.
\newblock Variation-stable fusion for ppg-based biometric system.
\newblock In {\em Proceedings of the International Conference on Acoustics,
  Speech and Signal Processing (ICASSP)}, pages 8042--8046. IEEE, 2021.

\bibitem{Hwang9130730}
Dae~Yon Hwang, Bilal Taha, Da~Saem Lee, and Dimitrios Hatzinakos.
\newblock Evaluation of the time stability and uniqueness in ppg-based
  biometric system.
\newblock {\em IEEE Transactions on Information Forensics and Security},
  16:116--130, 2021.

\bibitem{jain2004introduction}
Anil~K Jain, Arun Ross, and Salil Prabhakar.
\newblock An introduction to biometric recognition.
\newblock {\em IEEE Transactions on circuits and systems for video technology},
  14(1):4--20, 2004.

\bibitem{jain2021real}
Jawahar Jain, Vatche~A Attarian, Sajid Sadi, and Pranav Mistry.
\newblock Real time authentication based on blood flow parameters, July 2021.
\newblock US Patent 11,064,893.

\bibitem{karimian2019attack}
Nima Karimian.
\newblock How to attack ppg biometric using adversarial machine learning.
\newblock In {\em Proceedings of the Autonomous Systems: Sensors, Processing,
  and Security for Vehicles and Infrastructure}, volume 11009, page 1100909.
  International Society for Optics and Photonics, 2019.

\bibitem{karimian2017human}
Nima Karimian, Zimu Guo, Mark Tehranipoor, and Domenic Forte.
\newblock Human recognition from photoplethysmography (ppg) based on
  non-fiducial features.
\newblock In {\em Proceedings of the International Conference on Acoustics,
  Speech and Signal Processing (ICASSP)}, pages 4636--4640. IEEE, 2017.

\bibitem{RESITKAVSAOGLU20141}
A.~Reşit Kavsaoğlu, Kemal Polat, and M.~Recep Bozkurt.
\newblock A novel feature ranking algorithm for biometric recognition with ppg
  signals.
\newblock {\em Computers in Biology and Medicine}, 49:1--14, 2014.

\bibitem{kolkur2016human}
S Kolkur, D Kalbande, P Shimpi, C Bapat, and J Jatakia.
\newblock Human skin detection using rgb, hsv and ycbcr color models.
\newblock In {\em Proceedings of the International Conference on Communication
  and Signal Processing}, pages 324--332. Atlantis Press, 2016.

\bibitem{lewandowska2011measuring}
Magdalena Lewandowska, Jacek Rumi{\'n}ski, Tomasz Kocejko, and Jedrzej Nowak.
\newblock Measuring pulse rate with a webcam --- a non-contact method for
  evaluating cardiac activity.
\newblock In {\em Proceedings of the Federated Conference on Computer Science
  and Information Systems}, pages 405--410. IEEE, 2011.

\bibitem{NEURIPS2020_e1228be4}
Xin Liu, Josh Fromm, Shwetak Patel, and Daniel McDuff.
\newblock Multi-task temporal shift attention networks for on-device
  contactless vitals measurement.
\newblock In {\em Proceedings of the Advances in Neural Information Processing
  Systems}, volume~33, pages 19400--19411. Curran Associates, Inc., 2020.

\bibitem{Lovisotto9150630}
Giulio Lovisotto, Henry Turner, Simon Eberz, and Ivan Martinovic.
\newblock Seeing red: Ppg biometrics using smartphone cameras.
\newblock In {\em Proceedings of the IEEE/CVF Conference on Computer Vision and
  Pattern Recognition Workshops}, pages 3565--3574, virtual, 2020. IEEE.

\bibitem{lu2021dual}
Hao Lu, Hu Han, and S~Kevin Zhou.
\newblock Dual-gan: Joint bvp and noise modeling for remote physiological
  measurement.
\newblock In {\em Proceedings of the IEEE/CVF Conference on Computer Vision and
  Pattern Recognition}, pages 12404--12413, 2021.

\bibitem{GPflow2017}
Alexander G. de~G. Matthews, Mark {van der Wilk}, Tom Nickson, Keisuke. Fujii,
  Alexis {Boukouvalas}, Pablo {Le{\'o}n-Villagr{\'a}}, Zoubin Ghahramani, and
  James Hensman.
\newblock {{GP}flow: A Gaussian process library using TensorFlow}.
\newblock {\em Journal of Machine Learning Research}, 18(40):1--6, apr 2017.

\bibitem{Meziatisabour9346017}
Rita Meziatisabour, Yannick Benezeth, Pierre De~Oliveira, Julien Chappe, and
  Fan Yang.
\newblock Ubfc-phys: A multimodal database for psychophysiological studies of
  social stress.
\newblock {\em IEEE Transactions on Affective Computing}, 2021,
  doi:10.1109/TAFFC.2021.3056960.

\bibitem{mocco2018new}
Andreia~V Mo{\c{c}}o, Sander Stuijk, and Gerard de Haan.
\newblock New insights into the origin of remote ppg signals in visible light
  and infrared.
\newblock {\em Scientific Reports}, 8(1):1--15, 2018.

\bibitem{oung2020live}
Stephen Oung, Avrum~Douglas Hollinger, Gregor Simeonov, and Abhishek Ranjan.
\newblock Live user authentication device, system and method, Oct. 2020.
\newblock US Patent App. 16/956,470.

\bibitem{pedregosa2011scikit}
Fabian Pedregosa, Ga{\"e}l Varoquaux, Alexandre Gramfort, Vincent Michel,
  Bertrand Thirion, Olivier Grisel, Mathieu Blondel, Peter Prettenhofer, Ron
  Weiss, Vincent Dubourg, et~al.
\newblock Scikit-learn: Machine learning in python.
\newblock {\em The Journal of Machine Learning Research}, 12:2825--2830, 2011.

\bibitem{poh2010non}
Ming-Zher Poh, Daniel~J McDuff, and Rosalind~W Picard.
\newblock Non-contact, automated cardiac pulse measurements using video imaging
  and blind source separation.
\newblock {\em Optics Express}, 18(10):10762--10774, 2010.

\bibitem{Rathore3410158}
Aditya~Singh Rathore, Zhengxiong Li, Weijin Zhu, Zhanpeng Jin, and Wenyao Xu.
\newblock A survey on heart biometrics.
\newblock {\em ACM Computing Surveys}, 53(6):1--38, Dec. 2020.

\bibitem{schafer2011savitzky}
Ronald~W Schafer.
\newblock What is a savitzky-golay filter?
\newblock {\em IEEE Signal processing magazine}, 28(4):111--117, 2011.

\bibitem{Seepers7892894}
Robert~Mark Seepers, Wenjin Wang, Gerard de Haan, Ioannis Sourdis, and Christos
  Strydis.
\newblock Attacks on heartbeat-based security using remote
  photoplethysmography.
\newblock {\em IEEE Journal of Biomedical and Health Informatics},
  22(3):714--721, 2018.

\bibitem{song2021pulsegan}
Rencheng Song, Huan Chen, Juan Cheng, Chang Li, Yu Liu, and Xun Chen.
\newblock Pulsegan: Learning to generate realistic pulse waveforms in remote
  photoplethysmography.
\newblock {\em IEEE Journal of Biomedical and Health Informatics},
  25(5):1373--1384, 2021.

\bibitem{verkruysse2008remote}
Wim Verkruysse, Lars~O Svaasand, and J~Stuart Nelson.
\newblock Remote plethysmographic imaging using ambient light.
\newblock {\em Optics Express}, 16(26):21434--21445, 2008.

\bibitem{wang2020brainprint}
Min Wang, Jiankun Hu, and Hussein~A Abbass.
\newblock Brainprint: Eeg biometric identification based on analyzing brain
  connectivity graphs.
\newblock {\em Pattern Recognition}, 105:107381, 2020.

\bibitem{wang2015novel}
Wenjin Wang, Sander Stuijk, and Gerard De~Haan.
\newblock A novel algorithm for remote photoplethysmography: Spatial subspace
  rotation.
\newblock {\em IEEE Transactions on Biomedical Engineering}, 63(9):1974--1984,
  2015.

\bibitem{wang2021attacks}
Xuerui Wang, Zheng Yan, Rui Zhang, and Peng Zhang.
\newblock Attacks and defenses in user authentication systems: A survey.
\newblock {\em Journal of Network and Computer Applications}, 188:103080, 2021.

\bibitem{Yadav8411233}
Umang Yadav, Sherif~N. Abbas, and Dimitrios Hatzinakos.
\newblock Evaluation of ppg biometrics for authentication in different states.
\newblock In {\em Proceedings of the International Conference on Biometrics},
  pages 277--282, 2018.

\bibitem{yu2019remote}
Zitong Yu, Wei Peng, Xiaobai Li, Xiaopeng Hong, and Guoying Zhao.
\newblock Remote heart rate measurement from highly compressed facial videos:
  an end-to-end deep learning solution with video enhancement.
\newblock In {\em Proceedings of the IEEE/CVF International Conference on
  Computer Vision}, pages 151--160, 2019.

\bibitem{zhang2021h2k}
Junqing Zhang, Yushi Zheng, Weitao Xu, and Yingying Chen.
\newblock H2k: A heartbeat-based key generation framework for ecg and ppg
  signals.
\newblock {\em IEEE Transactions on Mobile Computing}, 2021,
  doi:10.1109/TMC.2021.3096384.

\bibitem{zhang2016joint}
Kaipeng Zhang, Zhanpeng Zhang, Zhifeng Li, and Yu Qiao.
\newblock Joint face detection and alignment using multitask cascaded
  convolutional networks.
\newblock {\em IEEE Signal Processing Letters}, 23(10):1499--1503, 2016.

\bibitem{Zhao9155526}
Tianming Zhao, Yan Wang, Jian Liu, Yingying Chen, Jerry Cheng, and Jiadi Yu.
\newblock Trueheart: Continuous authentication on wrist-worn wearables using
  ppg-based biometrics.
\newblock In {\em Proceedings of the Annual IEEE International Conference on
  Computer Communications (INFOCOM)}, pages 30--39, 2020.

\bibitem{zhao2020trueheart}
Tianming Zhao, Yan Wang, Jian Liu, Yingying Chen, Jerry Cheng, and Jiadi Yu.
\newblock Trueheart: Continuous authentication on wrist-worn wearables using
  ppg-based biometrics.
\newblock In {\em Proceedings of the IEEE INFOCOM 2020-IEEE Conference on
  Computer Communications}, pages 30--39. IEEE, 2020.

\end{thebibliography}

}

\end{document}